\documentclass[twocolumn, tighten, resetfootnote]{aastex701}
\usepackage{amsmath}

\newcommand{\cosmic}{\textsc{cosmic}}
\newcommand{\compas}{\textsc{compas}}
\newcommand{\cosmerge}{\textsc{cosmerge}}
\newcommand{\tng}{\textsc{tng100}}

\begin{document}

\title{A Stellar Role Reversal: Multiple Features in the Mass and Mass Ratio Distributions of Merging Binary Black Holes from Stable Mass Transfer}

\author[orcid=0009-0003-5399-3798]{Gina Chen}
\affiliation{McWilliams Center for Cosmology and Astrophysics, Department of Physics, Carnegie Mellon University, Pittsburgh, PA 15213, USA}
\email{gwchen@andrew.cmu.edu}

\author[orcid=0000-0001-5228-6598]{Katelyn Breivik}
\affiliation{McWilliams Center for Cosmology and Astrophysics, Department of Physics, Carnegie Mellon University, Pittsburgh, PA 15213, USA}
\email{kbreivik@andrew.cmu.edu}

\author[orcid=0000-0001-5484-4987]{Lieke van Son}
\affiliation{Department of Astrophysics/IMAPP, Radboud University Nijmegen, PO Box 9010, 6500 GL Nijmegen, The Netherlands}
\email{lieke.vanson@ru.nl}

\correspondingauthor{G. Chen}
\email{gwchen@andrew.cmu.edu}

\begin{abstract}

Observations of gravitational wave events have enabled the measurement of the merging binary black hole (BBH) mass function. This mass function encodes the physical interactions which shape the formation and evolution of BBHs. In this work we investigate how the stable mass transfer (SMT) channel of BBH formation imprints onto the BBH primary mass and mass ratio distributions. We use both an analytic framework and binary population synthesis to show how assumptions about mass transfer accretion efficiency and mass transfer stability affect the BBH mass distribution. Under the assumption of conservative mass transfer, we find that the SMT channel produces two observationally distinct subpopulations: a high primary mass, near equal mass ratio population formed through mass ratio reversal (MRR), and a low primary mass non-MRR subpopulation. The mass range where MRR occurs is determined by assumptions about binary SMT. In particular, we find that the stability criteria for mass transfer at different stellar evolutionary stages carve out complementary regions in the primary-mass--mass-ratio plane, separating the MRR and non-MRR populations into distinct peaks at high and low primary mass respectively. Our results imply that the physics of SMT creates distinct features in gravitational wave populations which current and near future gravitational wave detectors may be able to resolve.

\end{abstract}

\section{Introduction} \label{sec:intro}

Gravitational waves were first detected from the merger of a binary black hole (BBH) just over a decade ago \citep{abbot2016}. Since this discovery, the catalog of observed BBH mergers has grown to contain more than 130 events \citep{gwtc1, gwtc2, gwtc3, gwtc4, gwtc5}. A wide range of masses have been discovered, with features in the mass distribution at $10 M_\odot$ and $35 M_\odot$, and features in the mass ratio distribution are beginning to be resolved \citep{gwtc3_population, gwtc4_population, gwtc5_population}. Despite this, the origins of these BBHs are still uncertain. 

Many channels have been proposed for BBH formation, which can be broadly categorized into isolated formation, where the BBH progenitor forms and evolves without external influences, and dynamical formation, where gravitational interactions or accretion disk dynamics can facilitate the merger of the BBH \citep[see e.g.,][]{mandel2022, breivik2026}. Several studies have suggested that the observed population of BBH mergers contains multiple subpopulations which could indicate contributions from different formation channels \citep{wong2021, stevenson2022, godfrey2023, banagiri25, ray2025, ray2026, flanagan2026}. However, the relative rates of each channel are difficult to ascertain due to large uncertainties in stellar evolution, binary interactions, and external influences, as well as degeneracies between the measurable properties of BBHs formed across different channels \citep{zevin21, cheng2023}. Investigating each formation channel in detail across several parameters helps build understanding about which features of the BBH population formed are distinct to that channel and how each assumption influences those properties \citep{loc_feat, NPWV}. 

The stellar components in wide binaries are expected to evolve without interacting with their companion, except in the case of external interactions with the galactic tide \citep{stegmann2024} or flybys \citep[e.g.,][]{michaely2019}, though the BBH merger rate from this channel is sensitive to assumptions for natal kicks imparted to BHs upon formation \citep[e.g.,][]{raveh2022}. At the closest separations, chemically homogeneous evolution of massive stars can arise due to tidal spinup in very short orbits, preventing the expansion of each stellar component, such that a BBH can form and merge within a Hubble time \citep[e.g.,][]{marchant2016, demink2016}. Most massive stars form in binaries where the stellar components interact through Roche-lobe overflow \citep[RLOF,][]{sana12}. The outcome of this RLOF can remain stable or become dynamically unstable based on the response of the stellar components or the binary orbit as mass is exchanged \cite[e.g.,][]{hjellming87, ge2015, klencki2021, marchant2021}. In this work we focus on the stable mass transfer only (SMT-only) channel of BBH formation, in which all RLOF phases during a binary's evolution remain stable \citep[e.g.,][]{vandenHeuvel2017, inayoshi2017, vanSon2020, gallegosgarcia21, bavera2021,olejak2021, picco2024, klencki2026}. This excludes the classic CE channel, in which the first mass transfer is stable but the second is unstable \citep[e.g.][]{broekgaarden22a}.

A key uncertainty in the SMT-only channel is the mass transfer accretion efficiency. We denote this efficiency $\beta$, defined as the fraction of mass lost by the donor star that is accreted by its companion, the accretor. Observations of post mass transfer systems can provide constraints on the accretion efficiency \citep[e.g.,][]{villasenor2023, bao2025, vanSon2026}. A common assumption is that the accreting star rapidly spins up to critical rotation \citep{packet1981} and then ceases accretion \citep{paxton2015}. This yields highly non-conservative mass transfer for all wide systems in which tides are too weak to enforce synchronization. However, studies of observed post-interaction binaries span a wide range of inferred efficiencies from strongly non-conservative  \citep[e.g.,][]{petrovic2005, nuijten2025}, to accretion efficiencies greater than $50\%$ \citep[e.g.][]{lechien2025}. This motivates a broad agnostic exploration of how accretion efficiency shapes the resulting BBH population from the stable mass transfer channel.

This paper is structured as follows. In Section \ref{sec:simulation_description} we describe how we simulate a population of BBHs. We describe the typical SMT-only BBH formation channel in Section \ref{sec:smt-path}. In Section \ref{sec:two_peaks}, we show that the merging population of SMT-only BBHs can be clearly divided into two subpopulations across multiple observable characteristics. We discuss boundaries on these subpopulations, how those boundaries are affected by key binary evolution parameters, and how changing these parameters affects the BBH merger rate and mass distribution in Section \ref{sec:bse_params}. We discuss our results and summarize our conclusions in Section \ref{sec:conclusions}.

\section{Simulating a BBH merger population} \label{sec:simulation_description}

We use the binary population synthesis code \cosmic\ to construct a merging BBH population observable by gravitational waves. First we simulate a catalog of BBHs assuming a single-age burst of star formation on a grid of metallicities. Then we apply a cosmic star formation history (SFH) which is based on fits to the IllustrisTNG simulation to construct a cosmological population of merging BBHs. In this section, we describe each of these steps in detail. 

\subsection{Initializing a starburst population with \cosmic} \label{subsec:cosmic}

\cosmic\footnote{\href{https://cosmic-popsynth.github.io}{cosmic-popsynth.github.io}} is an open-source, Python-based binary population synthesis code that applies fitting formulae to single star evolution simulations following \citet{hurley2000} and the binary evolution algorithm of \citet{bse} with updated treatment of binary interactions and massive star evolution. We refer the reader to \citet{cosmic} for the comprehensive details of these updates, but include descriptions of assumptions that are critical to the formation and evolution of BBH populations below. We use \href{https://github.com/COSMIC-PopSynth/COSMIC/releases/tag/v4.1.0}{\cosmic\ V4.1.0} for all simulations in this work.

We first consider a fiducial binary evolution model with the following stellar evolution and binary interaction assumptions. Massive stars lose mass via metallicity-dependent winds according to the prescriptions defined in \citet{vink2001} and \citet{vink2005}, which are fit to observations of hydrogen-rich OB and Wolf-Rayet stars respectively. In all cases, if a star has a luminosity beyond the Humphreys-Davidson limit of $L>6\times10^{5}\,L_{\odot}$ we apply a luminosity-dependent mass loss rate according to \cite{hurley2000}. This causes massive stars to more rapidly strip their envelopes and evolve toward hotter temperatures, with the aim of mimicking the effect of luminous blue variable eruptions.

When a binary enters RLOF, we apply several assumptions for how mass transfer proceeds. The mass transfer stability criteria is based on the donor star's evolutionary state and the binary mass ratio \citep[e.g., ][]{hjellming87,soberman1997}. Main sequence stars (Case A donors) are assumed to have $\zeta_\mathrm{ad} = 2$, Hertzsprung gap stars (Case B) have $\zeta_\mathrm{ad} = 6.5$, and core helium burning stars (Case C) have $\zeta_\mathrm{ad} = 4.75$, where $\zeta_\mathrm{ad} = d\ln R / d\ln M$ is the adiabatic mass-radius exponent. This translates to critical mass ratios of $M_\mathrm{donor} / M_\mathrm{accretor} = 1.717$, $3.825$, and $3$, respectively, and this combination of mass ratio assumptions is denoted \texttt{qcflag=5} in the \cosmic\ documentation. 

During RLOF the donor star loses mass at a rate proportional to its Roche-filling factor defined by \citet{bse}. For our fiducial model, we assume fully conservative mass transfer ($\beta = 1$) for stellar accretors and ten times the Eddington limit for mass transfer onto compact objects \citep[e.g.,][]{begelman2002}. Any mass lost from the system during Roche-overflow mass transfer carries away angular momentum as if it is a wind from the accretor. Prior to the onset of RLOF, the binary is assumed to undergo tidal circularization and synchronization that persists through the duration of RLOF.
 
BH masses are assigned based on stellar progenitor properties at core collapse according to the ``delayed'' supernova explosion prescription from \citet{fryer}. We note that this prescription does not differentiate between the formation of neutron stars and BHs except for a defined maximum neutron star mass (or minimum BH mass) which we assume to be $3\,M_{\odot}$. For each BH formed, we draw a natal kick from a lognormal distribution with $\mu = 5.6$ and $\sigma = 0.69$, based on the \citet{disberg2025} pulsar catalog. This kick is then reduced by the fraction of mass that falls back onto the compact object as determined by our compact object formation prescription. As a general rule of thumb, natal kicks are fully suppressed by fallback for BHs that form with masses above $15M_\odot$.

When a star undergoes pair instability supernova (PISN), we apply a polynomial fit to the single star models described in \citet{marchant19} (see Table 1) to define the minimum core mass at which PISN occurs. With this prescription, PISN causes a pileup of BHs at high masses, with a maximum BH mass of $44 M_\odot$.

We simulate BBH populations for a grid of 30 metallicities evenly spaced in $\log_{10}Z$, ranging from $10^{-4}$ to $0.03$. Initial masses are drawn from a \citet{kroupa01} initial mass function that ranges from $0.08-100 M_\odot$. We additionally assume a binary fraction of 0.7, meaning each star drawn from the initial mass function has a 70\% chance of having a companion star. Companion star masses are sampled from a uniform mass ratio distribution, where the minimum allowed mass ratio is defined such that the pre-MS lifetime of the companion star is shorter than the MS lifetime of the primary star. For massive stars, this applies a general lower mass ratio limit of $q \sim 0.1$. Orbital periods and eccentricities are drawn from the power law distributions described in \citet{sana12}, which have exponents $\pi = -0.55$ and $\eta = -0.45$, respectively, and are based on observations of O stars in nearby stellar clusters. The minimum period is either $10^{0.15}$ days or the minimum period to prevent RLOF, and the maximum period is $10^{5.5}$ days.

\cosmic\ continues to sample and evolve binaries until the population distribution of user-specified properties satisfies the convergence limit supplied. We use the component masses, separation, and eccentricity distributions at BBH formation and set the convergence limit to $-5$. The convergence limit leads to adaptively sized populations depending on how the binary interaction assumptions imprint into the orbital properties of the merging BBH population.  For a detailed description of the convergence process, see \cite{cosmic}. We restrict our analysis to the population of merging BBHs for better comparison to data from ground-based gravitational wave detectors. Broadly, increased metallicity leads to a reduction of the formation of BBH mergers per unit solar mass. At low metallicities, our simulations result in approximately to $10^4$ BBH mergers, formed from an initialized stellar mass of $\sim 10^9 M_\odot$. Near solar metallicity, the formation rate reduces to fewer than $10$ from over $10^{10} M_\odot$ in initialized stellar mass.

\subsection{Assumed star formation history} \label{subsec:sfh}

The SFH describes the rate of star formation at a given redshift and metallicity. We adopt the functional form proposed by \citet{M&D}, which consists of a cosmic star formation rate density component that is independent of metallicity and a metallicity distribution component that depends on redshift. We use the SFH from \citet{loc_feat}, which is fit to the \tng\ simulation of the IllustrisTNG suite \citep{illustris1, illustris2, illustris3, illustris4, illustris5} and uses a skewed log-normal distribution for the metallicity distribution component. 

\subsection{Applying an SFH with \cosmerge} \label{subsec:cosmerge}

\cosmerge\footnote{\href{https://github.com/COSMIC-PopSynth/cosmerge}{github.com/COSMIC-PopSynth/cosmerge}} is a publicly available, open-source post-processing suite that is designed to read in \cosmic\ data and apply a user-defined SFH to create a catalog of BBH mergers. With the set of \cosmic\ catalogs described in Section \ref{subsec:cosmic} as the ``pool'' of BBH mergers to draw from and the SFH model described in Section \ref{subsec:sfh}, we sample a new catalog of $10^6$ BBH mergers that accounts for metallicity-specific star formation for redshifts $z<10$. We then filter this catalog to the population of SMT-only binaries that merge within a Hubble time.

To calculate the cosmological merger rate, we apply a normalization factor that divides the integrated mass formed across redshift in a cubic Gpc by the average mass sampled per BBH merger, weighted by the metallicity distribution of our SFH. This factor is denoted \texttt{norm\_fac} in \cosmerge.  

\section{The stable mass transfer channel for BBH mergers} \label{sec:smt-path}

\begin{figure}[ht]
    \centering
    \includegraphics[width=0.9\columnwidth]{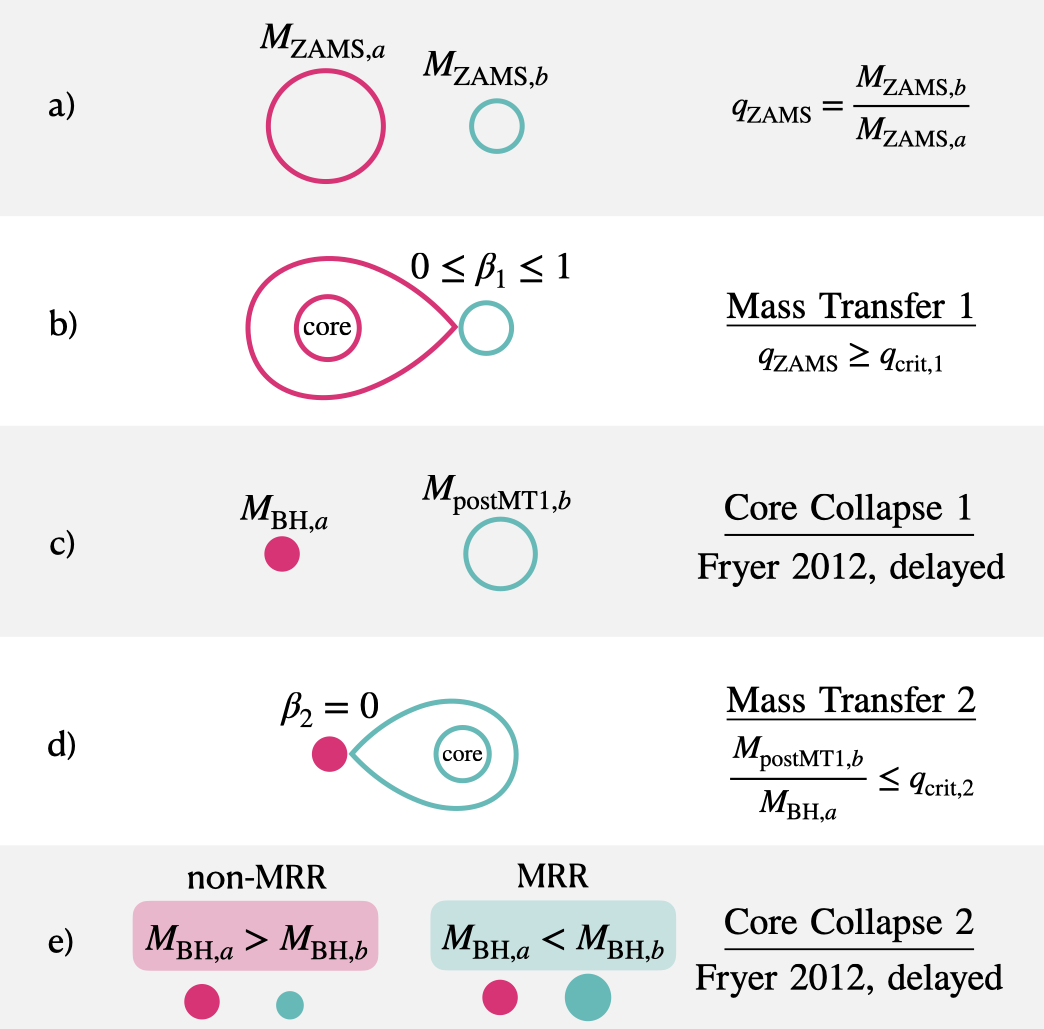} 
    \caption{A schematic showing the major evolutionary stages in the stable mass transfer (SMT) only channel.}
    \label{fig:smtchannel}
\end{figure} 

Figure \ref{fig:smtchannel} shows the key evolutionary stages typical of the SMT-only channel of BBH formation. The initially more massive star begins the first stage of RLOF mass transfer and goes through core collapse before the initially less massive star fills its Roche lobe and also eventually reaches core collapse. Both stages of mass transfer are subject to stability criteria and both instances of core collapse can produce natal kicks. 80\% of the merging BBHs in our \cosmerge\ simulation follow these stages of evolution. An additional 15\% of the merging population experiences three phases of RLOF, where the first instance of RLOF occurs when the more massive star is on the main sequence, resulting in a small amount of mass transfer \citep[Case AB mass transfer e.g.,][]{wellstein2001}, and the second and third phases of RLOF proceed as described in the figure.

We define the index $a$ such that it refers to the star or black hole that evolves from the more massive star at ZAMS and define the index $b$ to refer to the star or black hole that evolves from the less massive star at ZAMS. Unless stated otherwise, we define the mass ratio $q$ as $M_b / M_a$ for each stage of binary evolution. By definition, $M_{\mathrm{ZAMS},a} \geq M_{\mathrm{ZAMS},b}$, and therefore, the mass ratio at ZAMS, $q_\mathrm{ZAMS} = M_{\mathrm{ZAMS},b} / M_{\mathrm{ZAMS},a}$, is always between 0 and 1. We will use the phrase ``BBH primary" or simply ``primary" to refer to the more massive component of the BBH, and the phrase ``BBH secondary" or ``secondary" to refer to the less massive component of the BBH. Additionally, an index of $1$ will be used to refer to values relevant to the first mass transfer event (MT1, stage b of Figure \ref{fig:smtchannel}) and an index of $2$ to the second (MT2, stage d of Figure \ref{fig:smtchannel}). We note that in \cosmic, the critical mass ratio is always defined as donor mass divided by accretor mass, which is the inverse of $q_{\mathrm{crit},1}$, as star $b$ is the accretor during MT1. 

The stability of a phase of mass transfer is determined by the envelope structure of the donor star and the orbital response to the mass transfer. We adopt the naming convention of Case A, Case B, and Case C mass transfer to refer to mass transfer with main sequence, Hertzsprung gap, and core helium burning donors at the onset of RLOF. We additionally define Case BC mass transfer as mass transfer that begins while the donor is on the Hertzsprung gap and transitions to core helium burning during RLOF.

During MT2, accretion onto BH $a$ is limited to ten times the Eddington factor, such that $\beta_2 \approx 0$. We assume that the non-accreted mass carries away angular momentum as if it is a wind from BH $a$. Star $b$ loses its envelope over the course of MT2, eventually reaches core collapse, and forms a BH with mass $M_{\rm{BH,b}}$. In the case where $M_{\mathrm{BH},b} > M_{\mathrm{BH}, a}$, we define the BBH as having undergone a mass ratio reversal (MRR; see blue population in Figure \ref{fig:obs}). 

\section{Results}

\subsection{The stable mass transfer channel produces distinct features in the BBH merger mass and mass-ratio plane} \label{sec:two_peaks}

\begin{figure*}[ht] 
    \centering
    \includegraphics[width=0.8\textwidth]{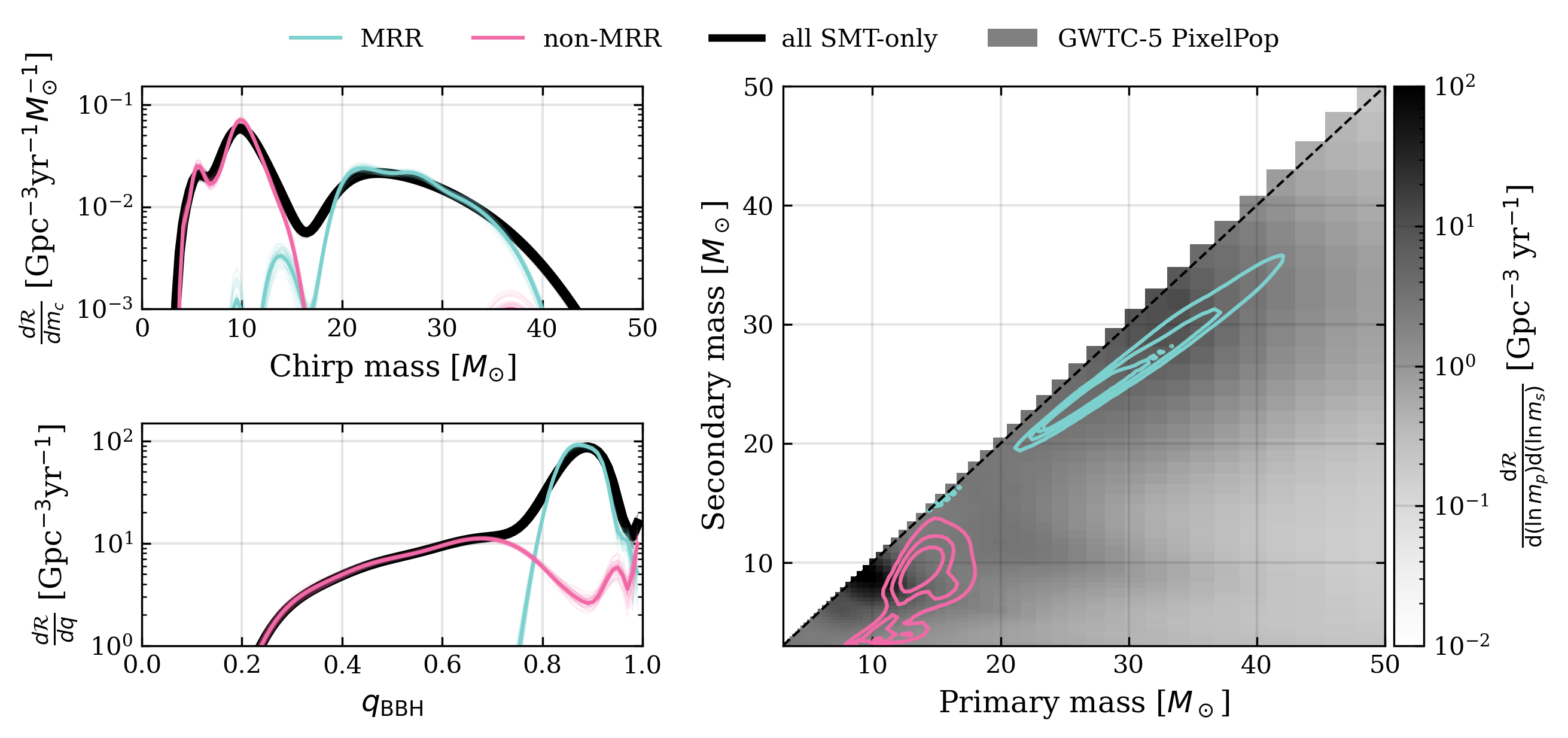} 
    \caption{Merger rates as a function of the chirp mass (top left), BBH mass ratio (bottom left), and individual component masses (right) for the merging BBH population evaluated at $z=0.1$. Black solid lines denote the entire SMT-only population, while pink and blue lines denote the non-MRR and MRR subpopulations respectively. The gray histogram on the right panel is the GTWC-5 component mass distribution, using the PixelPop model. \textit{Top left:} The SMT-only population produces two peaks, corresponding to low mass non-MRR BBHs and high mass MRR BBHs, with the split occurring at chirp masses around $15 M_\odot$. \textit{Bottom left:} All MRR BBHs have $q_\mathrm{BBH} > 0.75$, while the non-MRR distribution ranges from $0.2$ to $1$ and has peaks near $0.4$ and $0.7$ \textit{Right:} The MRR subpopulation roughly overlaps with the observed high-mass GW sources, while the non-MRR subpopulation partially overlaps with observed low-mass GW sources.} 
    \label{fig:obs}
\end{figure*}

BBH component masses are the primary observables for ground-based gravitational wave surveys. In this subsection we compare our simulated SMT-only BBH mergers to the GWTC-5 population.
In Figure \ref{fig:obs} we plot the chirp mass distribution, BBH mass ratio distribution, and joint component mass distribution of our fiducial SMT-only, merging BBH population. We define the BBH mass ratio $q_\mathrm{BBH}$ as the mass of the BBH secondary divided by the mass of the BBH primary, as this is what is observable by gravitational wave detectors. 

In the left two panels, we show the marginalized kernel density estimates (KDEs) for the chirp mass and mass-ratios of the merging BBH population, evaluated at $z=0.1$. We additionally regenerate each KDE $50$ times on a bootstrapped random sample with replacement of the same size to determine how sensitive the distribution is to random statistical noise arising from our sample size. Each resampled KDE is shown as a thin line for each subpopulation. On the right panel of Figure \ref{fig:obs} we plot marginalized two-dimensional BBH component mass KDE evaluated at $z=0.1$. We show the 25th, 50th, and 75th percentiles of the MRR subpopulation in blue and non-MRR subpopulation in pink. We show the observed population from GWTC-5 using the PixelPop model from \citet{gwtc5_population} in grey, where the axes have been limited to a minimum of $3 M_\odot$ in order to show only the BBH population consistent with our minimum BH mass assumption.

The MRR and non-MRR subpopulations form clearly distinct features in all three panels. We find a low chirp mass feature that peaks near $10 M_\odot$ which is formed entirely from the non-MRR systems, and a broad high mass secondary feature peaking at $20 M_\odot$ and spanning the range $20-40 M_\odot$ which consists entirely of MRR systems. This is consistent with recent results by \citet{godfrey2026}, who show that the low-mass peak in the GWTC-4 catalog is likely formed entirely by non-MRR BBHs. The all SMT-only KDE does not perfectly trace the MRR KDE near $40 M_\odot$ because a small number of non-MRR BBHs are formed with high masses due to PISN, similar to \citet{smith2026}. The non-MRR subpopulation contains BBHs with a broad mass ratio distribution ranging from $q_\mathrm{BBH} \sim 0.3$ to $0.7$. The MRR subpopulation contains BBHs with mass ratios strongly peaked around $q_\mathrm{BBH} \sim 0.9$, with less than 0.5\% of MRR BBHs forming with mass ratios below $q_\mathrm{BBH} \sim 0.8$. These results are consistent with previous studies of MRR populations \citep[e.g.,][]{broekgaarden22a, smith2026}. The MRR BBH mergers in our simulation align well with the higher mass BBH mergers observed by GWTC-5, while the non-MRR BBH mergers in our simulations are distributed over a wider range of mass ratios than the observed population and are limited to a more narrow range of primary masses. We note that a simple primary mass cut at $20 M_\odot$ effectively separates the MRR and non-MRR subpopulations.  

\subsection{The location of SMT features is driven by mass transfer stability and accretion efficiency} \label{sec:bse_params}

The features in the chirp mass function and component mass distributions in Figure \ref{fig:obs} are correlated with differences in the first mass transfer phase. All MRR binaries with primary masses less than $20 M_\odot$ undergo Case B or C mass transfer, while 86\% of MRR BBHs with primary masses greater than $20 M_\odot$ go through Case A mass transfer. For the non-MRR binaries, 80\% of BBHs with secondary masses less than $8 M_\odot$ experience Case C mass transfer, while only 5\% of non-MRR binaries with secondary masses greater than $8 M_\odot$ undergo Case C mass transfer. 

Our simulations show a strong correlation between MRR and features in the BBH merger mass distributions. Specifically, the minimum allowed primary mass increases with increasing $q_\mathrm{ZAMS}$, with boundaries on $q_\mathrm{ZAMS}$ set by the mass transfer stability for different mass transfer cases. We use the analytical model from \citet{NPWV} to understand the behavior of this relationship and to help us understand the characteristics of the non-reversed and mass ratio reversed populations. 

\begin{figure*}[ht] 
    \centering
    \includegraphics[width=0.8\textwidth]{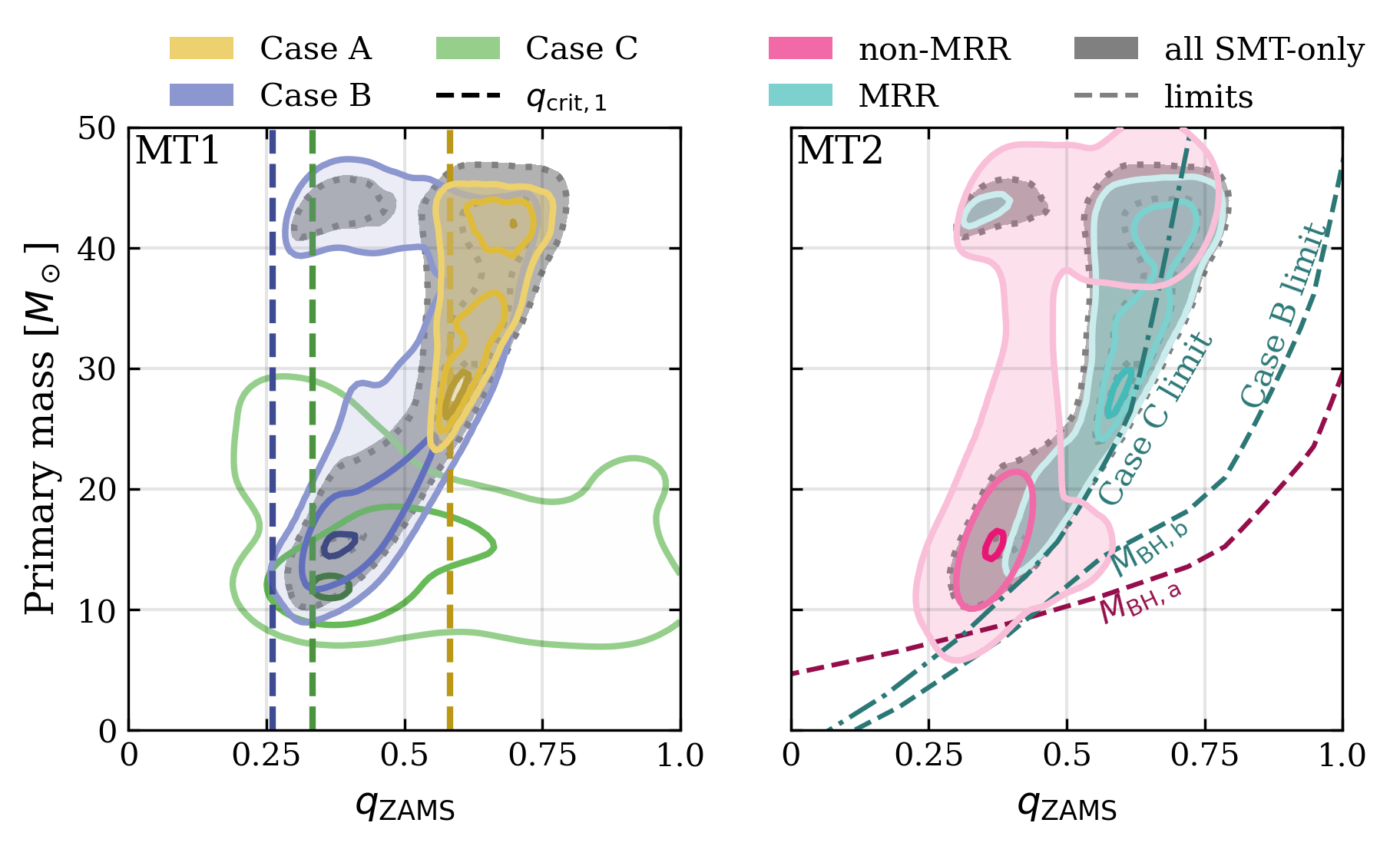} 
    \caption{The primary masses and initial mass ratios of the merging SMT-only BBHs. We normalize each population separately, and use the opacity to indicate the relative size of each subpopulation. The gray dotted contours and shading show the full SMT-only population.  \textit{Left panel:} BBHs are classified by the donor star's evolutionary state during MT1. Vertical lines denote $q_{\mathrm{crit},1}$ for each type of donor star, which is approximately the minimum allowed $q_\mathrm{ZAMS}$ for each subpopulation. \textit{Right panel:} BBHs are classified by whether they are MRR or non-MRR. The dashed pink line indicates the expected minimum mass of BH $a$, while the dashed and dot-dashed blue lines indicate the expected minimum mass of BH $b$ depending on star $b$'s evolutionary stage during MT2.} 
    \label{fig:outlines}
\end{figure*}

\citet{NPWV} describe an analytic model that connect $\beta_1$, $q_{\mathrm{crit},1}$, and $q_{\mathrm{crit},2}$ to minimum component BH masses as a function of $q_\mathrm{ZAMS}$. We outline this model and discuss the key modifications we made in order to match \cosmic\ simulations more faithfully. For a full derivation, we refer the reader to Appendix \ref{app:derivation}, where we detail limiting assumptions for each stage shown in Figure \ref{fig:smtchannel}. 

We assume that binaries start with $q_\mathrm{ZAMS} \leq q_{\mathrm{crit}, 1}$ in order to fulfill the stability requirement for MT1. $\beta_1$ determines how much of the envelope of star $a$ is accreted onto star $b$ during MT1 (stage b of Figure \ref{fig:smtchannel}). While \citet{NPWV} applies a fixed core mass fraction that agrees with their \compas\ simulations, we introduce a core-mass fraction that varies with mass. We approximate this $f_\mathrm{core}$ with the piecewise function

\begin{equation} \label{eq:varfcore}
    f_\mathrm{core} = 0.41 + 
        \begin{cases} 
          a_\mathrm{core} (M_\ast - M_\mathrm{turn}) & M_\ast < M_\mathrm{turn} \\
          b_\mathrm{core} (M_\ast - M_\mathrm{turn}) & M_\ast \geq M_\mathrm{turn} \\
       \end{cases}
\end{equation}

\noindent where $a_\mathrm{core} = 0.0018$, $b_\mathrm{core} = 0.00074$, $M_\mathrm{turn} = 58 M_\odot$, and $M_\ast$ is the mass of the star at the start of RLOF. The exact value of $f_\mathrm{core}$ for a specific star can vary depending on both the stage of stellar evolution and the time when mass transfer is initiated; this function approximates the upper limit of the core to envelope mass ratio in our \cosmic\ simulations. We choose the upper limit because this sets the lowest possible analytic minimum (see right panel of Figure \ref{fig:variations}). 

We calculate supernova mass loss with the piecewise function

\begin{equation} \label{eq:cc_mass_loss}
    dM_\mathrm{CC}(M_\mathrm{core}) = 
        \begin{cases} 
          a_\mathrm{CC} M_\mathrm{core} + b_\mathrm{CC} & M_\mathrm{core} \leq M_\mathrm{thresh} \\
          0.5 M_\odot & M_\mathrm{core} > M_\mathrm{thresh} \\
       \end{cases}
\end{equation}

\noindent where $a_\mathrm{CC} = -0.9$, $b_\mathrm{CC} = 13.9$, and $M_\mathrm{thresh} = 14.8 M_\odot$. For core masses above $M_\mathrm{thresh}$, we assume full fallback, but include a $0.5 M_\odot$ mass loss term to account for neutrino mass loss. When we calculate the minimum BH masses, we enforce the piecewise function threshold, resulting in a more stringent minimum mass limit than the one shown in \citet{NPWV}.

With these approximations, we can calculate the expected mass of BH $a$. We additionally assume $\beta_2$ is fixed to $0$ for MT2 (stage d of Figure \ref{fig:smtchannel}) to account for Eddington-limited accretion. With this, we can calculate the mass of BH $b$ as well. Under our fiducial assumptions, whether a binary becomes MRR depends on $q_\mathrm{ZAMS}$, while the stability criteria define mass limits for SMT-only evolution.

In Figure \ref{fig:outlines}, we plot the entire merging SMT-only BBH population's primary masses and initial mass ratios and compare to assumptions in our analytic model. On the left panel we split the population by donor star type during MT1, where yellow, blue, and green indicate Case A, B, and C mass transfer, respectively. On the right panel, we split the population into the non-MRR and MRR subpopulations, which are outlined in pink and blue. In both panels, the dotted gray contour indicates the full SMT-only merging population. We use a two-dimensional KDE of the primary masses and initial mass ratios to determine the 5th, 50th, and 95th percentiles of each subpopulation. The shading for each subpopulation is determined by the fraction of the full merging SMT-only population that the subpopulation makes up.

We normalize each KDE separately, so the contours in Figure \ref{fig:outlines} are not indicative of the relative rates of each subpopulation. We report the relative rates here for clarity and completeness. About $68\%$ of merging SMT-only BBHs undergo Case A mass transfer during MT1, while $30\%$ of the population experiences Case B mass transfer. Only $2\%$ of the total population goes through Case C mass transfer during MT1. $68\%$ and $32\%$ of the SMT-only merging BBHs are MRR and non-MRR, respectively.

The two main features visible in the chirp mass panel of Figure \ref{fig:obs} are shown clearly here: a high mass MRR subpopulation and a low mass non-MRR subpopulation. The bimodality of the MRR BBH subpopulation shown in the left two panels of Figure \ref{fig:obs} is visible here as well: a low BBH primary mass population which forms from binaries with $q_\mathrm{ZAMS} \lesssim 0.6$ and a high BBH primary mass population where $q_\mathrm{ZAMS} \gtrsim 0.6$. 

The vertical dashed lines on the left panel show the stability criteria for each mass transfer type, while the dashed lines on the right panel indicate the calculated analytic minima for each BH component mass as a function of the initial mass ratio. The pink and blue dashed lines are the analytically derived minimum masses for BH $a$ and $b$, respectively, assuming that the BBHs undergo Case B mass transfer during MT2. This is true for 75\% of the non-MRR subpopulation, and the analytic minimum derived is a good approximation. However, the MRR subpopulation more closely follows the analytically derived minimum mass of BH $b$ using the $q_{\mathrm{crit}, 2}$ value that corresponds to Case C mass transfer, shown in Figure \ref{fig:outlines} as a blue dot-dashed line. This is because every MRR BBH in our model experiences Case BC mass transfer during MT2. Case C mass transfer has a more stringent stability criteria, so the MRR subpopulation is better described by the Case C line, despite initiating mass transfer while on the Hertzsprung gap. 

Case A mass transfer has the most stringent stability criteria in our fiducial model, so the BBHs that undergo Case A mass transfer during MT1 are limited to the highest $q_\mathrm{ZAMS}$ values, as shown on the left panel. A higher $q_\mathrm{ZAMS}$ limit also increases the minimum possible primary mass for that subpopulation, as shown on the right panel. Additionally, because donors on the main sequence have not fully formed their convective core yet, they typically have a slightly smaller $f_\mathrm{core}$, resulting in a larger amount of envelope available to donate and enhancing MRR. Therefore, BBHs that undergo Case A mass transfer during MT1 form the majority of the high-mass MRR population. Despite these binaries starting out very close together, they undergo significant orbital widening over the course of MT1 because we assume $\beta_1 = 1$ in our fiducial model. This widening allows star $b$ to evolve to nearly the end of the Hertzsprung gap phase before overfilling its Roche lobe and initiating MT2. 

The low primary mass non-MRR binaries near $\sim 10-15M_\odot$ spanning $q_\mathrm{ZAMS} \sim 0.2-0.9$ are formed by binaries that experience Case C mass transfer during MT1. These do not follow the analytic minima because they start MT1 late in star $a$'s evolution and undergo core collapse during RLOF, interrupting the mass transfer. Star $b$ does not accrete the full amount of mass expected by our analytic model during MT1, allowing these binaries to remain stable during MT2. The Case C subpopulation extends further past the analytically assumed $q_\mathrm{ZAMS} = q_\mathrm{crit}$ limit than the other subpopulations, because BBHs that undergo Case C mass transfer during MT1 have the most time to evolve before RLOF begins and thus have a longer time to lose mass via winds.

There is a small island of MRR BBHs at $q_\mathrm{ZAMS} \sim 0.3$ and primary masses above $40 M_\odot$, as well as a pileup of non-MRR BBHs with similar primary masses, both of which are caused by PISN. These are also visible in the chirp mass and component mass panels of Figure \ref{fig:obs}.

The location and occurrence of MRR and non-MRR features in the mass and mass ratio distribution are due to boundaries set by mass transfer stability for different mass transfer cases. The combination of the lower $q_\mathrm{ZAMS}$ limit, determined by $q_{\mathrm{crit}, 1}$ and the minimum primary mass as a function of $q_\mathrm{ZAMS}$, determined by the accretion efficiency $\beta_1$ and $q_{\mathrm{crit},2}$, set strong limits on the allowed primary masses for each subpopulation of SMT-only BBHs. These limits can be mapped directly onto different features of the primary mass distribution.

\subsection{More efficient accretion increases the fraction and masses of mass ratio reversed binaries} \label{sec:rates}

\begin{figure}[ht] 
    \centering
    \includegraphics[width=0.75\columnwidth]{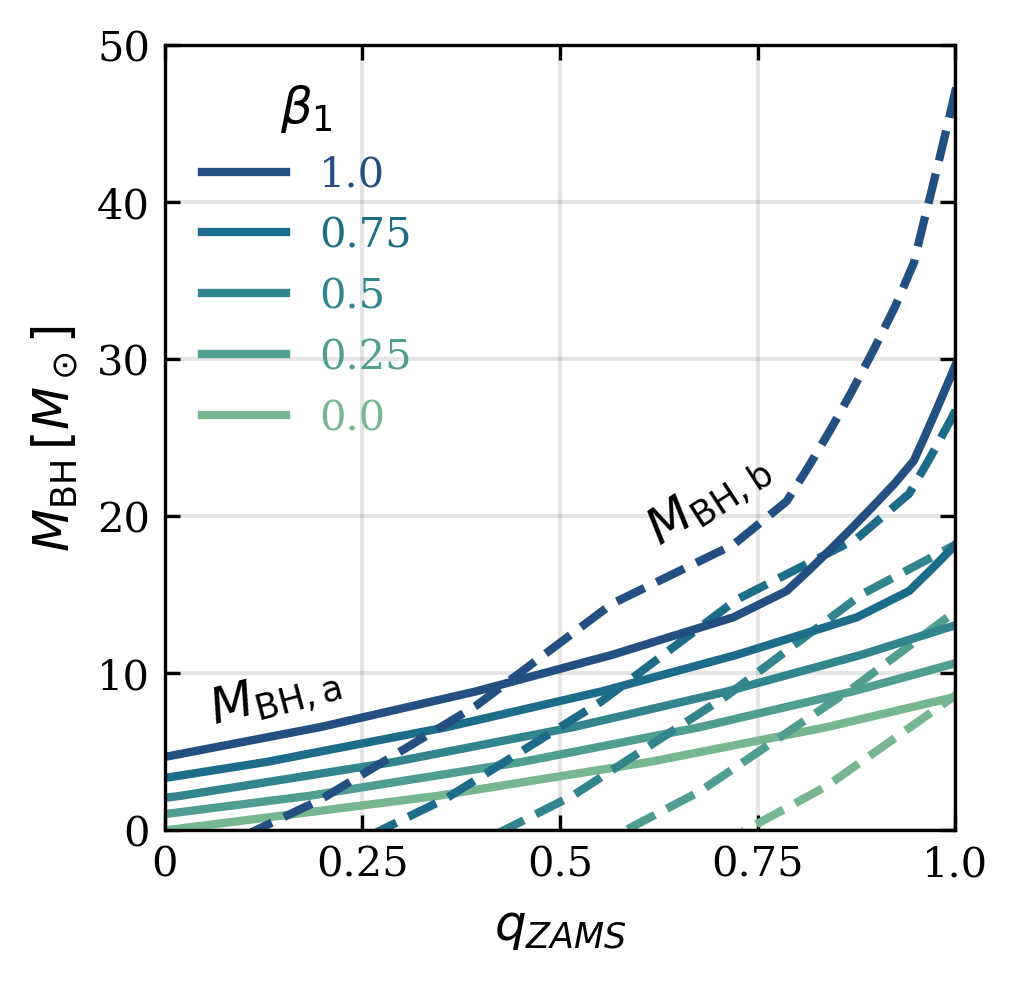} 
    \caption{The minimum mass of each component of a BBH formed through the SMT-only channel according to our analytic model for the five $\beta_1$ values we simulated. Increasing $\beta_1$ results in a higher minimum limit for both $M_{\mathrm{BH},a}$ and $M_{\mathrm{BH}, b}$} 
    \label{fig:analytic_beta_var}
\end{figure}

\begin{figure*}[ht] 
    \centering
    \includegraphics[width=0.95\textwidth]{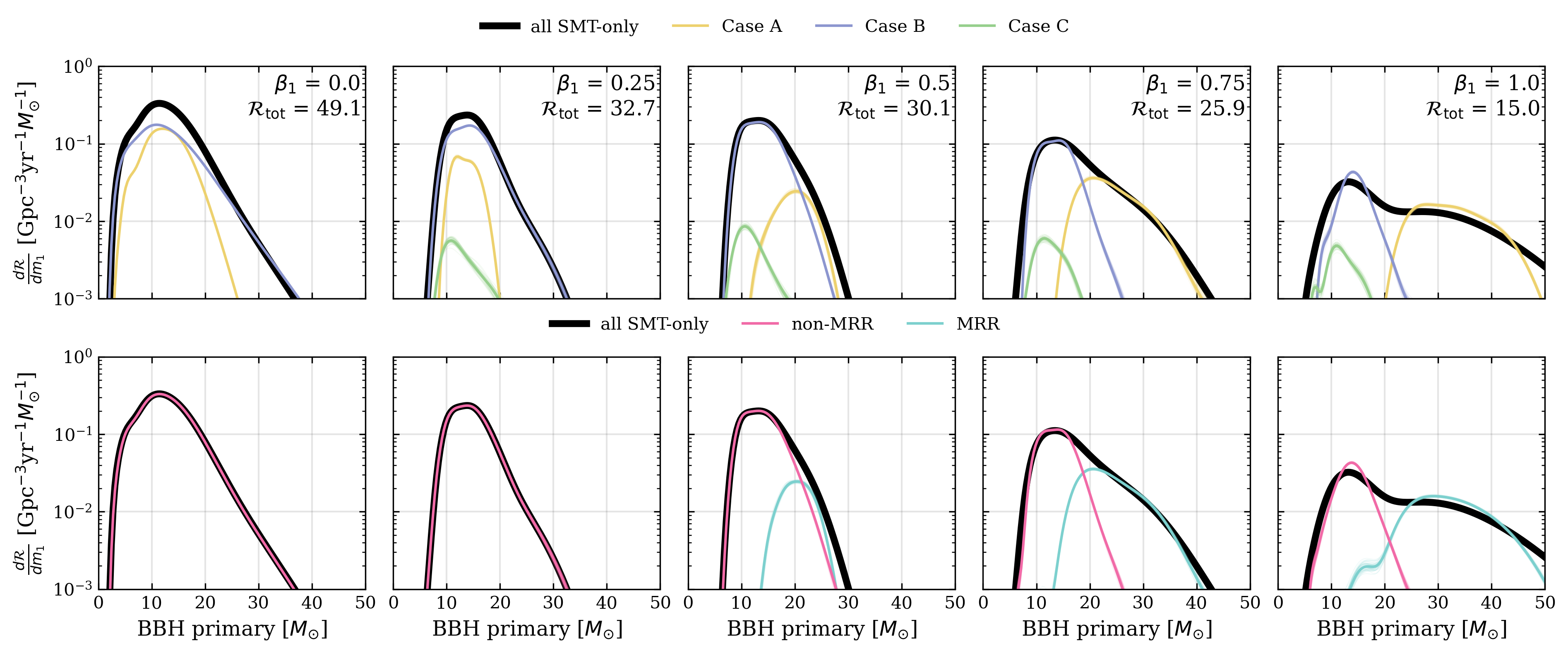} 
    \caption{The merger rate as a function of primary mass for different $\beta_1$ values. In the top row, the population is split into MRR (blue) and non-MRR (pink) subpopulations, and in the bottom row, the population is split by donor star type during MT1. The full SMT-only merger rate $\mathcal{R}_\mathrm{tot}$ evaluated at $z=0.1$ is given in units of $\mathrm{Gpc}^{-3}\mathrm{yr}^{-1}$. Black lines denote the full SMT-only population in both rows. We do not include the subpopulations when they make up less than $1\%$ of the total merging population. \textit{Top row:} At low $\beta_1$, the binaries that undergo Case A and Case B mass transfer during MT1 both peak at $\sim 10 M_\odot$. The Case C subpopulation shifts toward higher primary masses as $\beta_1$ increases, while the Case B subpopulation remains relatively stationary. \textit{Bottom row:} The non-MRR subpopulation peaks at low primary masses ($< 20M_\odot$) across all $\beta_1$ variations, while the MRR subpopulation shifts to higher primary masses and widens with increasing $\beta_1$. } 
    \label{fig:rates}
\end{figure*}

In order to determine the effect of $\beta_1$ on the merger rate and mass and $q_\mathrm{BBH}$ distributions, we simulate BBHs at an additional four fixed $\beta_1$ values ($0$, $0.25$, $0.5$, and $0.75$) and compare to our fiducial model, where $\beta_1 = 1.0$. Mass and angular momentum lost during non-conservative mass transfer is treated as a wind from the accretor. All other binary evolution assumptions and population construction methods are the same as the fiducial model described in Section \ref{sec:simulation_description}.

In Figure \ref{fig:analytic_beta_var}, we plot the minimum expected mass of BH $a$ and BH $b$ in solid and dashed lines, respectively, according to our analytic model (described in Section \ref{sec:bse_params}; full derivation in Appendix \ref{app:derivation}) for each of the five $\beta_1$ values we simulate. We assume $q_{\mathrm{crit}, 2} = 3.825$ for this plot, which is our fiducial assumption for Case B mass transfer. Increasing $\beta_1$ increases the minimum allowed mass for BH $b$, as star $b$ accretes a larger fraction of the mass lost by star $a$ during MT1. Since $q_{\mathrm{crit}, 2}$ is fixed, MT2 is more likely to become unstable as the mass of star $b$ grows relative to the mass of BH $a$. This increases the minimum allowed mass of $M_{\mathrm{BH},a}$. Therefore, a higher $\beta_1$ value limits the SMT-only channel toward higher mass BHs, consistent with the findings in \citet{NPWV}. 

\begin{figure*}[ht] 
    \centering
    \includegraphics[width=0.95\textwidth]{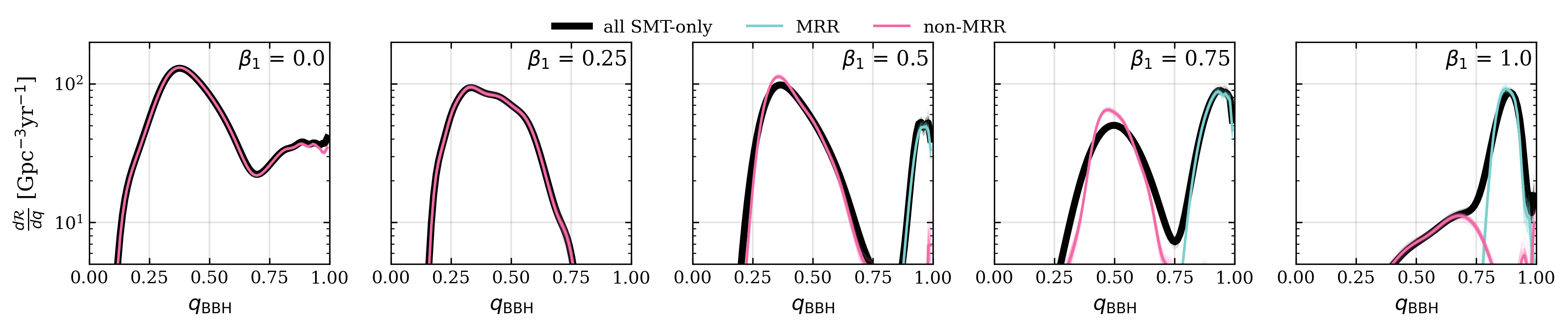} 
    \caption{The merger rate as a function of BBH mass ratio for different $\beta_1$ values. As in Figure \ref{fig:rates}, black lines show the distribution for the entire SMT-only BBH population, while blue and pink lines show the distributions for the MRR and non-MRR subpopulations, respectively. The MRR subpopulation is consistently limited to $q_\mathrm{BBH} > 0.75$, while the non-MRR subpopulation spans a wide range of lower $q_\mathrm{BBH}$ values.} 
    \label{fig:rates_q}
\end{figure*}

In Figure \ref{fig:rates} we plot the BBH merger rate for each $\beta_1$ variation as a function of primary mass. In black, we show the entire merging SMT-only BBH population, and the top row is split by the evolutionary stage of the donor during MT1 and the bottom row is split by whether the BBH is MRR. We use the same KDE and bootstrapping method as described in Section \ref{sec:two_peaks} for all panels. The merger rate for each $\beta_1$ variation is shown on each panel in units of $\mathrm{Gpc}^{-3} \mathrm{yr}^{-1}$. As $\beta_1$ increases, the merger rate decreases because increasing $\beta_1$ results in a smaller allowed parameter space for binaries to undergo stable mass transfer, as discussed for Figure \ref{fig:analytic_beta_var}. As $\beta_1$ decreases, the prevalence of the MRR subpopulation decreases significantly. We only plot it when it accounts for at least $1\%$ of entire merging SMT-only population, which occurs when $\beta_1 \geq 0.5$ in our simulations. Similarly, we do not plot the subpopulation that undergoes Case C mass transfer during MT1 in the $\beta_1 = 0$ panel because it accounts for less than $1\%$ of the overall population. 

The high-mass, MRR peak widens considerably and shifts toward higher masses at higher $\beta_1$ values. It peaks between $15-30 M_\odot$ at $\beta_1 = 0.5$, widens to $\sim 15 - 40 M_\odot$ at $\beta_1=0.75$, and spans $\sim 15-45 M_\odot$ at $\beta_1=1.0$. We note the extended tail out to $50 M_\odot$ is a result of the KDE, and the maximum $BH$ mass in our simulations is limited to $45 M_\odot$ due to PISN assumptions (see Section \ref{subsec:cosmic}). The fraction of mergers that are MRR increases with increasing $\beta_1$, and these BBHs predominantly undergo Case A mass transfer during MT1. Because Case A mass transfer has the most stringent stability criteria, the only binaries that remain stable through MT1 must necessarily be the higher mass systems, because at lower masses the high accretion efficiency results in unstable mass transfer during MT2. 

The location of the low-mass, non-MRR peak remains constant near $10 M_\odot$, narrowing slightly as $\beta_1$ increases. The allowed range of $q_\mathrm{ZAMS}$ that forms non-MRR BBHs narrows with increasing accretion efficiency, and the higher $q_\mathrm{ZAMS}$ (and thus higher mass) binaries become more likely to form MRR BBHs. When $\beta_1 \geq 0.5$, the non-MRR peak becomes dominated by binaries that undergo Case B mass transfer during MT1 because the Case B stability criteria allows much lower mass binaries to remain stable through MT2.

Shifts in the primary mass distribution across different $\beta_1$ values correspond to different initial binary populations. This is because changing $\beta_1$ affects not only the masses and thus the stability of the second stage of mass transfer, but also the orbital separation after MT1. Only the fiducial model, where $\beta_1 = 1.0$, has a dip at $20 M_\odot$ (see Figure \ref{fig:obs}) because MT1 occurs predominantly as Case B mass transfer at lower $\beta_1$ values ($82\%$ and $52\%$ of the entire SMT-only population for $\beta_1=0.5$ and $\beta_1 = 0.75$, respectively), as opposed to only $29\%$ of the SMT-only population for the fiducial model, $\beta_1=1.0$. In the fiducial model, a larger fraction ($68\%$) of merging SMT-only binaries undergo Case A mass transfer during MT1 because increased orbital widening during MT1 increases the separation at which BBHs form. Additionally, binaries that undergo Case A mass transfer and remain stable must necessarily be higher mass due to the higher $q_\mathrm{ZAMS}$ limit enforced by the stability criteria for Case A mass transfer.

In Figure \ref{fig:rates_q}, we plot the merger rate as a function of BBH mass ratio for all $\beta_1$ variations. We again plot the entire SMT-only population in black and the non-MRR and MRR subpopulations in pink and blue, respectively. We use the same KDE method described in Section \ref{sec:two_peaks}. 

In order for MRR BBHs to form, their mass ratio after MT1 must be asymmetric enough that the binary remains MRR after envelope stripping during MT2. The merging MRR subpopulation thus consistently forms only at high $q_\mathrm{BBH}$ across $\beta_1 > 0.5$. The minimum $q_\mathrm{BBH}$ is driven by mass transfer stability during MT2. Binaries with the most extreme mass ratio reversals (corresponding to lower $q_\mathrm{BBH}$) form from systems that begin with $q_\mathrm{ZAMS}$ near 1. For $\beta > 0.5$, these binaries then also have extreme mass ratios at the onset of MT2, making them more likely to become unstable (see middle panel of Figure \ref{fig:variations}). Because less mass is accreted onto star $b$ during MT1 when $\beta \simeq 0.5$, the systems that become MRR are those that just barely made it over the threshold and are thus close to equal in mass. For $\beta_1 > 0.5$, enough mass is accreted during MT1 that more extreme MRR is possible, leading to a wider range of $q_\mathrm{BBH}$. Because conservative mass transfer leads to significant orbital widening, longer delay times limit the number of merging BBHs at low $q_\mathrm{BBH}$ for $\beta_1=1.0$.

The non-MRR subpopulation, on the other hand, must form from the lower $q_\mathrm{ZAMS}$ binaries. The peak of the $q_\mathrm{BBH}$ feature shifts to higher values with increasing $\beta_1$ (from $q_\mathrm{BBH} \sim 0.3$ for $\beta_1=0.0$ to $q_\mathrm{BBH} \sim 0.7$ at $\beta_1 = 1.0$), which is expected as star $b$ gains increasing fractions of the mass lost by star $a$ during MT1. Additionally, at low $\beta_1$, the $q_\mathrm{BBH}$ distribution is shifted toward lower values because the BBHs that are formed are typically quite low mass, meaning a lot of the secondaries, which are under $15 M_\odot$, will have lost mass during partial-fallback core collapse.

\section{Discussion} \label{sec:discussion}

\subsection{Caveats} \label{subsec:caveats}

\textit{Stellar radii:} We use the \citet{hurley2000} stellar evolution tracks as a basis for determining stellar radii, which are known to produce overinflated stars, especially at masses beyond $50 M_\odot$ \citep[e.g.][]{agrawal2020}. \cosmic\ is currently under active development to include updated stellar evolution tracks with improved radius estimates using the METhods of Interpolation for Single Star Evolution (\texttt{METISSE}) code package \citep{agrawal2023,agrawal2025}. Modifications to single star evolution could significantly change the evolutionary pathway, since different radial evolution will affect the properties of the binary at the onset of RLOF. For a detailed description of implementation of the \texttt{COSMIC-METISSE} framework and the effect of using updated single star models on the merging BBH population, see \citet{maclean2026:inprep}.

\textit{Wind mass loss:} Our analytic model assumes no wind mass loss between any of the stages of binary evolution, but wind mass loss is included in our \cosmic\ simulations. Winds are partially accounted for in our variable $f_\mathrm{core}$ assumption, which results in larger core masses relative to the fixed core masses used in the analytic model of \citet{NPWV}. Much higher wind mass loss rates on the hydrogen or helium main sequence could truncate core mass growth, leading to lower core masses and therefore BH masses \citep{bavera2023}.

\textit{Fixed mass transfer stability and accretion efficiency:} The use of a single-valued $q_\mathrm{crit}$ for each stage of stellar evolution is a simplification common to many binary population synthesis codes. \citet{gallegosgarcia21} show that this assumption, as implemented in \cosmic, overpredicts the occurrence of common envelope events compared to equivalent \texttt{MESA} models. While \cosmic\ does not currently implement stability criteria which account for donor or accretor response to mass loss or gain, we have shown how changes in $q_\mathrm{crit}$ affect the bounds of the resulting BBH population. These effects can be used to understand how a more complex stability criteria would impact the SMT-only channel. 

We additionally assume a fixed value of $\beta_1$ in all of our simulations for simplicity. \citet{nuijten2025} investigate a catalog of observed post-interaction WR+O binaries and find that the upper limit for $\beta_1$ differed between the likely Case A and likely Case B populations. The likely Case A binaries prefer low accretion efficiencies, and the likely case B binaries allow for accretion efficiencies as high as $\beta_1 = 1$. Our model shows how changing values of $\beta_1$ and $q_{\mathrm{crit}, 1}$ interact to affect the minimum allowed BH masses produced through the SMT-only channel.

\textit{Compact object formation:} Compact object mass prescriptions are uncertain, and newer proposed prescriptions are available in \cosmic\ \citep[e.g.][]{mandel2020, maltsev2025}. Our analytic model was based on the work of \citet{NPWV}, whose supernova mass loss function was also based on the \citet{fryer} prescription. For this reason, and for easier comparison with other codes, we chose to use the same prescription. Applying a different remnant mass prescription, especially one without a monotonic BH mass formation criteria, will produce a significantly different BBH mass function \citep[e.g.,][]{willcox2025}. However, we note that assumptions for SMT and mass transfer stability will play a role in the merging BBH primary mass and mass ratio distribution, regardless of the compact object formation prescription.

\textit{BH spin:} We do not consider BH spin in this paper because the physical mechanism that produces BH spins is uncertain \citep[e.g.][]{qin2018, qin2019, fuller+ma2019}. Previous studies have proposed that tidal spinup of the second-formed BH in a binary could produce measurable spin for binaries with periods below $1$ day \citep{zevin+bavera2022}. In this case, our MRR subpopulation could have primaries with nonzero spins, while the non-MRR subpopulation will have secondaries with nonzero spin. \citet{broekgaarden22a} find exactly this trend in their simulations when using a spin prescription based on \citet{bavera2020}. However, several studies argue that there is more evidence for both BHs in a BBH to have nonzero spin, rather than a spinning and nonspinning companion \citep[e.g.,][]{mould2022, adamcewicz2024}. 

\citet{briel2026} find that for each binary in their simulations, the primary and secondary BH generally have similar spins which increase in magnitude with metallicity. This trend is due to wind mass loss rates which increase with increasing metallicity such that high spins result from initially close binaries at high metallicity. This effect is only possible in our models with high $\beta_1$, because in the case of low $\beta_1$, MT1 shrinks the binary such that short initial separations lead to stellar mergers.

\subsection{Related population synthesis work} \label{subsec:related_work}

\citet{broekgaarden22a} investigate MRR in the isolated binary evolution channel using \compas. They find that the fraction of MRR BBHs increases with $\beta_1$, and the majority of the MRR binaries are formed through the SMT-only channel. Additionally, the MRR BBHs form a distinct high-$q_\mathrm{BBH}$ and high primary mass feature compared to the non-MRR BBHs. These findings are broadly consistent with our simulations, although we note that \citet{broekgaarden22a} include both the SMT-only and CE channels in their analysis, whereas our work focuses only on the SMT-only channel. We leave a detailed comparison and investigation of the CE populations simulated by \cosmic\ to future work.

\citet{loc_feat} explore the SMT-only population using the population synthesis code \compas. They find a secondary high-mass feature at $\sim 20 M_\odot$, at a similar mass range to the MRR feature in our simulations. However, this secondary feature occurs at a lower mass in their models. This is likely due to the fact that once binaries initiate SMT in \compas, the evolution is assumed to finish mass transfer stably, rather than allowing a transition to CE. Therefore, low-mass binaries can undergo MT2 as fully Case B mass transfer, rather than transitioning to Case C and becoming unstable as shown in our simulations (see Figure \ref{fig:outlines}). This transition occurs frequently in our \cosmic\ simulations, and \citet{klencki2026} argue that this type of delayed dynamical instability places a lower separation limit on the SMT-only channel. \citet{temmink2025} identify possible observational signatures of future delayed dynamical instability, which they argue will be observable by Gaia. This transitionary channel warrants further investigation, but is out of the scope of this work. 

We note that the application of our simple analytic model adequately explains the behavior of the BBH population in two different population synthesis codes. While both codes apply the \citet{hurley2000} stellar fits, the evolution algorithms are completely independent; our similar results should thus inspire confidence in the application of SMT in each code. 

\citet{dorozsmai2024} investigate the impact of binary interaction parameters on BBH formation using the population synthesis code \texttt{SEBA}. They find that these parameters affect the relative rates of the SMT-only and CE channels, but together create BBH populations with similar mass distributions across parameter variations. Higher values of $\beta_1$ at fixed mass transfer stability and angular momentum loss assumptions increase the population of BBH mergers with primary masses greater than $20 M_\odot$.

\citet{briel2026} use \texttt{POSYDON}, a detailed population synthesis code, to investigate the bounds of the initial population which forms SMT-only BBHs. They find that the majority of BBHs in their simulation undergo Case A mass transfer during both MT1 and MT2. They argue that rapid population synthesis codes overestimate the prevalence of Case B mass transfer in the SMT-only channel. In our simulations, the MRR population is dominated by Case A mass transfer during MT1, while non-MRR is produced through Case B mass transfer. For MT2, both the MRR and non-MRR populations initiate Case B mass transfer. However, we note that due to fixed assumptions for mass transfer efficiency that depend on spinup of the accretor, highly efficient Case B mass transfer cannot be explored with \texttt{POSYDON}.

\section{Conclusions} \label{sec:conclusions}

In this paper, we investigate the SMT-only channel of BBH formation and show that it produces two observationally distinct subpopulations. In our fiducial model, the merging MRR subpopulation creates a high primary mass feature that spans $20-40 M_\odot$ and a narrow mass ratio feature where $q_\mathrm{BBH} > 0.8$. The non-MRR subpopulation creates a low primary mass feature that peaks near $10 M_\odot$ with a broad range of BBH mass ratios, peaking around $q_\mathrm{BBH} \sim 0.7$. 

We additionally provide a framework for understanding how the mass transfer accretion efficiency and mass transfer stability criteria interact to produce these features and influence each subpopulation. More efficient accretion leads to a lower SMT-only merger rate, a larger fraction of MRR BBHs, and higher primary masses in both the MRR and non-MRR subpopulations. More stringent mass transfer stability criteria limit the range of $q_\mathrm{ZAMS}$ at which binaries can remain in the SMT-only channel, which in turn restricts the MRR BBHs to higher primary masses. 

Finally, the evolutionary stage at which a star initiates RLOF shapes the features in the BBH mass and mass-ratio distributions. This is what causes the distinct split between the MRR and non-MRR subpopulations at $20 M_\odot$ in our fiducial model: the MRR population is primarily formed by binaries which undergo Case A mass transfer during MT1, while the non-MRR population is primarily formed by binaries which undergo Case B mass transfer. The much more stringent Case A mass transfer stability criteria limits the MRR population to binaries with high $q_\mathrm{ZAMS}$, and only those with sufficiently high initial masses remain stable during MT2. 

The features in the MRR and non-MRR subpopulations can also be explained by the evolutionary stage of star $a$ during MT1. In the MRR subpopulation, BBHs with primary masses less than $20 M_\odot$ undergo Case B or C mass transfer during MT1, while those with primary masses greater than $20 M_\odot$ undergo Case A mass transfer. In the non-MRR subpopulation, a large fraction of BBHs with secondary mass less than $8 M_\odot$ undergo Case C mass transfer, whereas the BBHs with secondary mass greater than $8 M_\odot$ typically undergo case A or B mass transfer. 

Future observation of features in the BBH mass and mass ratio plane may help us understand how SMT imprints in evolution of massive binary stars and the BBHs they produce. 

\section{Data availability}

All data and code necessary to recreate the figures for this work will be available on Zenodo upon publication.

\begin{acknowledgments}
 GC thanks the \cosmic\ research group at Carnegie Mellon University for helpful discussions and feedback. KB acknowledges support from the Falco-DeBenedetti Career Development Professorship. LvS is supported by VI.Veni.242.115, Grant ID \url{https://doi.org/10.61686/XVIAV86753}.
\end{acknowledgments}

\begin{contribution}

GC was responsible for carrying out all simulations and analysis and for writing and submitting the manuscript
KB oversaw the work, provided software support, and edited the manuscript. LvS devised the initial research concept, oversaw the work, and edited the manuscript.

\end{contribution}

\software{
\texttt{astropy} \citep{astropy:2013,astropy:2018,astropy:2022,astropy_11121433}, 
\texttt{Jupyter} \citep{2007CSE.....9c..21P,kluyver2016jupyter},
\texttt{matplotlib} \citep{Hunter:2007},
\texttt{numpy} \citep{numpy}, 
\texttt{pandas} \citep{mckinney-proc-scipy-2010,pandas_10304236},
\texttt{python} \citep{python},
\texttt{scipy} \citep{2020SciPy-NMeth,scipy_10155614}, 
\cosmic\ \citep{Breivik2020,COSMIC_20721229},
\texttt{legwork} \citep{LEGWORK_joss,LEGWORK_apjs,legwork_8436065},
\texttt{seaborn} \citep{Waskom2021}, 
\texttt{tqdm} \citep{tqdm_3551211}}

Software citation information aggregated using \texttt{\href{https://www.tomwagg.com/software-citation-station/}{The Software Citation Station}} \citep{software-citation-station-paper,software-citation-station-zenodo}.

\appendix
\restartappendixnumbering

\section{Derivation of analytic BH mass minima}\label{app:derivation}

For simplicity, we assume that the mass ratio at the first mass transfer is equal to $q_\mathrm{ZAMS}$. The stability condition for MT1 can therefore be written as $q_\mathrm{ZAMS} \geq q_{\mathrm{crit}, 1}$. We assume that star $a$ loses its entire envelope as a result of MT1, so that only the core of the star remains. The fraction of a star's mass that is contained within the core is denoted by $f_\mathrm{core}$. We estimate this fraction with Equation \ref{eq:varfcore}, which we repeat here for ease of reference:

\begin{equation}
    f_\mathrm{core} = 0.41 + 
        \begin{cases} 
          a_\mathrm{core} (M_\ast - M_\mathrm{turn}) & M_\ast < M_\mathrm{turn} \\
          b_\mathrm{core} (M_\ast - M_\mathrm{turn}) & M_\ast \geq M_\mathrm{turn} \\
       \end{cases}
\end{equation}

\noindent where $a_\mathrm{core} = 0.0018$, $b_\mathrm{core} = 0.00074$, $M_\mathrm{turn} = 58 M_\odot$, and $M_\ast$ is the mass of the star at the start of RLOF. Each star's envelope mass is then $M_\mathrm{env} = (1 - f_\mathrm{core})M_\star$, and we denote the core mass fractions of star $a$ and star $b$ with $f_{\mathrm{core},a}$ and $f_{\mathrm{core},b}$, respectively.

The mass of star $b$ after the first mass transfer event can be approximated with the equation

\begin{equation}\label{eq:post_mt1_b}
    M_{\mathrm{postMT1},b} = M_{\mathrm{ZAMS},b} + \beta_1 (1-f_{\mathrm{core},a})M_{\mathrm{ZAMS},a}.
\end{equation}

\noindent The mass of BH $a$ is approximately

\begin{equation}\label{eq:bh_a}
    M_\mathrm{BH, a} = f_{\mathrm{core},a}M_{\mathrm{ZAMS},a} - dM_{\mathrm{CC},a},
\end{equation}

\noindent where the second term in Equation \ref{eq:bh_a} is the mass lost during core collapse. We model mass loss during the formation of the BH with the parametric function

\begin{equation}\label{eq:dmsn}
    dM_\mathrm{CC}(M_\mathrm{core}) = 
        \begin{cases} 
          a_\mathrm{CC} M_\mathrm{core} + b_\mathrm{CC} & M_\mathrm{core} \leq M_\mathrm{thresh} \\
          0.5 M_\odot & M_\mathrm{core} > M_\mathrm{thresh} \\
       \end{cases}
\end{equation}

\noindent where $a_\mathrm{CC} = -0.9$, $b_\mathrm{CC} = 13.9$, and $M_\mathrm{thresh} = 14.8 M_\odot$\footnote{This equation and its associated constants are denoted with the subscript SN in \citet{NPWV}.}. This differs slightly from the equation given in \citet{NPWV}; we include a fixed $0.5 M_\odot$ of neutrino mass loss above the threshold core mass, based an upper cutoff for neutrino mass loss given by \citet{lattimer89}.

For the second mass transfer, because we assume Eddington-limited accretion onto BHs, we fix $\beta_2 = 0$ for simplicity. We additionally assume no mass loss between MT1 and MT2, so that Equation \ref{eq:post_mt1_b} describes the mass of star $b$ at the onset of MT2. The mass of BH $b$ can thus be approximately calculated analogously to the mass of BH $a$:

\begin{equation}\label{eq:bh_b}
    M_\mathrm{BH, b} = f_{\mathrm{core},b}M_{\mathrm{postMT1},b} - dM_{\mathrm{CC},b},
\end{equation}

\noindent where $M_{\mathrm{postMT1},b}$ is the stellar mass used to calculate $f_{\mathrm{core},b}$ and $dM_{\mathrm{CC},b}$. 

For a given $M_{\mathrm{ZAMS},a}$, we can calculate the maximum allowed $M_{\mathrm{ZAMS},b}$ using Equations \ref{eq:post_mt1_b}, \ref{eq:bh_a} and the stability criteria, $q_\mathrm{postMT1} \leq q_\mathrm{crit,2}$. This results in the following inequality:

\begin{equation}
    M_{\mathrm{ZAMS},b} \leq q_{\mathrm{crit},2} M_{\mathrm{BH}, a} + \beta_1 (1- f_{\mathrm{core},a})M_{\mathrm{ZAMS},a}.
\end{equation}

\noindent For a given $q_\mathrm{ZAMS}$, we can calculate a minimum pair of ZAMS masses and the resulting minimum BH component masses. We note that this differs from Equation 7 of \citet{NPWV} because we enforce the $dM_\mathrm{CC}$ threshold cutoff.

\begin{figure*}[ht] 
    \centering
    \includegraphics[width=0.85\textwidth]{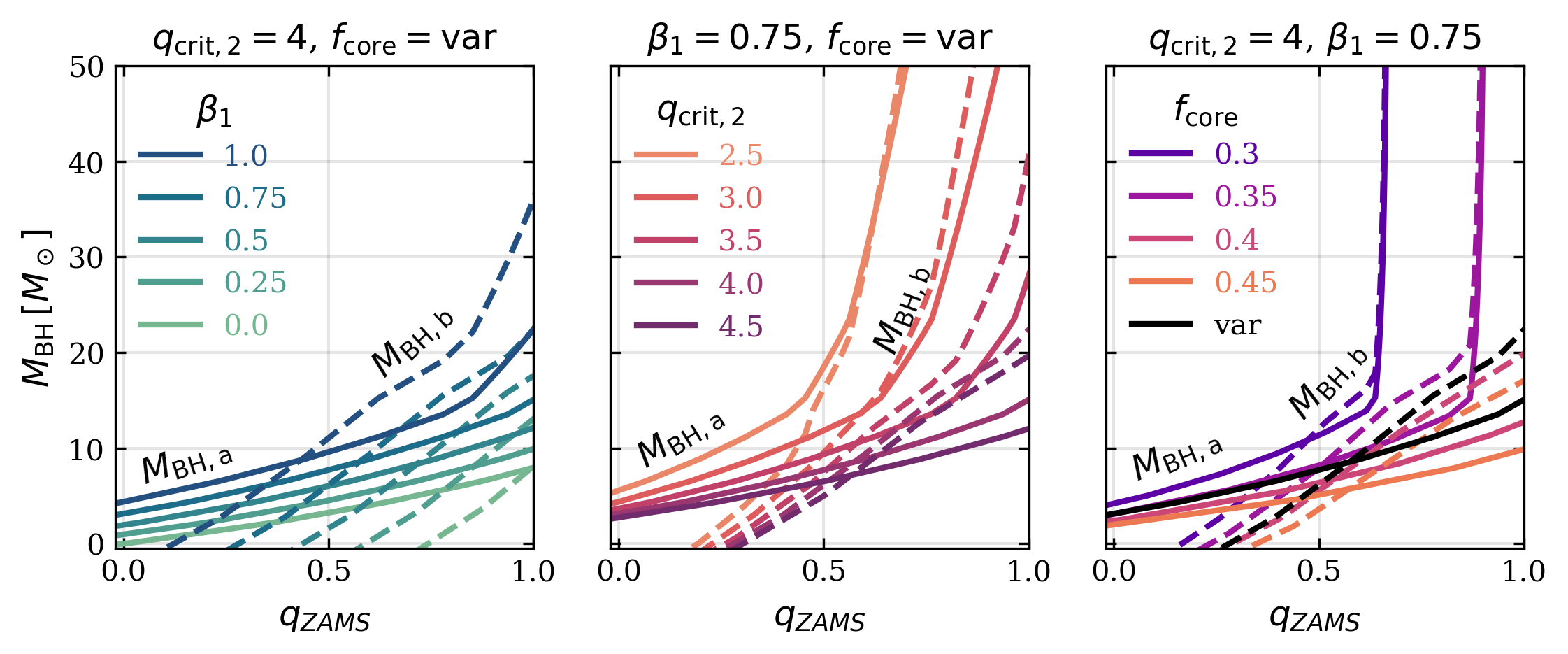} 
    \caption{The minimum possible $M_{\mathrm{BH}, a}$ (solid) and $M_{\mathrm{BH}, a}$ (dashed) for SMT-only BBHs plotted as a function the ZAMS mass ratio, $q_\mathrm{ZAMS}$. The panels, from left to right, show individual variations in $q_{\mathrm{crit},2}$, $\beta_1$, and $f_\mathrm{core}$, respectively. Otherwise, the parameters are fixed to $q_{\mathrm{crit},2} = 4$, $\beta_1 = 0.75$, and the variable $f_\mathrm{core}$ described by Equation \ref{eq:varfcore}. The ranges for each parameter are chosen to illustrate their impact, rather than to describe any specific model. All parameter combinations shown have a point at which the minimum mass of BH $b$ becomes larger than the minimum mass of BH $a$, indicating where MRR is likely. Increasing $\beta_1$ increases the minimum mass for each component and decreases the $q_\mathrm{ZAMS}$ at which the crossover occurs. Increasing $q_{\mathrm{crit},2}$ and $f_\mathrm{core}$ decrease the minimum masses and increase the crossover $q_\mathrm{ZAMS}$.}
    \label{fig:variations}
\end{figure*}

Figure \ref{fig:variations} shows the minimum BH masses calculated with varying values of $\beta_1$ on the left, varying values of $q_{\mathrm{crit,} 2}$ in the middle, and varying values of $f_\mathrm{core}$ on the right. Fixed values are set to $\beta_1 = 0.75$, $q_\mathrm{crit} = 4$, and the $f_\mathrm{core}$ described in Equation \ref{eq:varfcore} for easy comparison between panels. Note that this is not exactly the same as Figure \ref{fig:analytic_beta_var}, where $q_{\mathrm{crit},2} = 3.825$ to agree with our fiducial \cosmic\ model. In each panel, solid lines correspond to $M_{\mathrm{BH}, a}$, while the dashed lines correspond to $M_{\mathrm{BH},b}$. 

In the left panel, increasing the amount of accretion onto the secondary star leads to more massive systems in general. $M_\mathrm{BH, b}$ grows significantly larger with higher $\beta_1$, as more of the material from star $a$ is accreted. Because we assume $\beta_2 = 0$ in all cases, $\mathrm{min}(M_{\mathrm{BH}, a})$ is not affected as dramatically. However, a higher $\beta_1$ leads to larger $M_{\mathrm{postMT1},b}$ values, which increases the minimum required $M_\mathrm{BH, a}$ for the second mass transfer event to remain stable. Therefore, increasing $\beta_1$ leads to higher minimum masses for both BHs.

In the middle panel, we see the opposite trend, where increasing $q_\mathrm{crit, 2}$ allows fewer asymmetric mass ratio binaries to remain stable. Larger values of $q_{\mathrm{crit},2}$ mean smaller values of $M_{\mathrm{BH},a}$ are required for mass transfer to remain stable. Binaries with higher $q_\mathrm{ZAMS}$ will have larger $q_\mathrm{postMT1} \sim q_\mathrm{preMT2}$, so largest allowed $q_\mathrm{ZAMS}$ decreases with increasing $q_{\mathrm{crit},2}$. 

In the right panel, we show the effect of changing $f_\mathrm{core}$ on the minimum BBH masses. We additionally plot the minimum masses with the variable $f_\mathrm{core}$ described in Equation \ref{eq:varfcore} in black. This equation results in an $f_\mathrm{core}$ range of approximately $0.34$ to $0.42$, and thus it follows the $f_\mathrm{core} = 0.35$ lines fairly closely for most of the $q_\mathrm{ZAMS}$ range. For a fixed value of $f_\mathrm{core}$, our approximations create a nonphysical vertical cutoff due to $dM_\mathrm{CC}$ being a constant value at high stellar masses. Higher values of $f_\mathrm{core}$ push the minimum allowed BH mass down for both components, perhaps counterintuitively. For the same ZAMS masses, a larger $f_\mathrm{core}$ value leads to a larger $M_{\mathrm{BH},a}$ and smaller $M_{\mathrm{postMT1},b}$, resulting in a lower $M_{\mathrm{BH},b}$ and a lower minimum $M_{\mathrm{BH},b}$. For a fixed $M_{\mathrm{ZAMS}, b}$, systems with lower $M_{\mathrm{ZAMS},a}$ can remain stable, lowering the limit for $M_{\mathrm{BH},a}$ as well.

\restartappendixnumbering

\section{Mass transfer accretion efficiency variations}\label{app:full_pop_plots}

\begin{figure*}[ht] 
    \centering
    \includegraphics[width=0.99\textwidth]{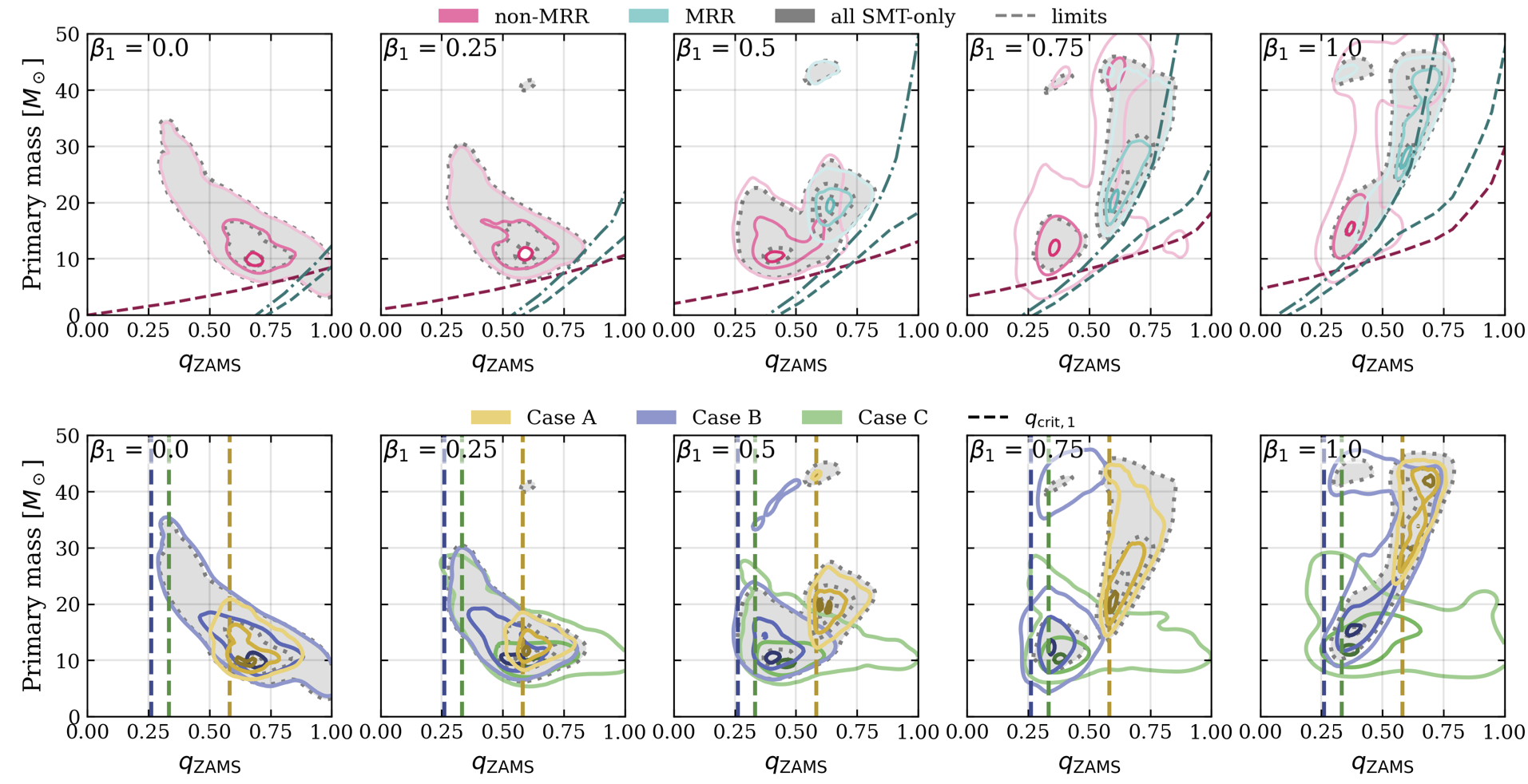} 
    \caption{Same as Figure \ref{fig:outlines}, for all $\beta_1$ variations explored in Section \ref{sec:rates}. Only the full SMT-only population has been shaded for clarity.}
    \label{fig:five_panel_plots}
\end{figure*}

In Figure \ref{fig:five_panel_plots}, we show the primary masses and initial mass ratios of the entire SMT-only population for each $\beta_1$ variation. We show the same subpopulations as Figure \ref{fig:outlines} for each simulation for reference. The MRR subpopulation overlaps with the subpopulation that undergoes Case A mass transfer during MT1. As $\beta_1$ increases, the minimum possible primary BH mass increases dramatically for the Case A subpopulation due to the interaction between the high $q_\mathrm{ZAMS}$ requirement for Case A mass transfer and the increasing minimum BH $b$ mass as a function of $q_\mathrm{ZAMS}$. The transition from Case B mass transfer to Case C mass transfer during MT2 is also apparent in all three $\beta_1$ variations with a significant MRR subpopulation, as shown by the dearth of BBHs between the Case B and Case C minimum lines in the top row.

\bibliography{bibliography}{}
\bibliographystyle{aasjournalv7}

\end{document}